\documentclass[aps,prl,twocolumn,groupedaddress,showpacs]{revtex4}

\usepackage{graphicx}
\begin{document}

\title{Matter X waves}
\author{Claudio Conti}
\email{c.conti@ele.uniroma3.it}
\homepage{http://optow.ele.uniroma3.it}
\author{Stefano Trillo}
\altaffiliation[Also with]{Dept. of Engineering,
University of Ferrara, Via Saragat 1, 44100 Ferrara, Italy}
\affiliation{
National Institute for the Physics of Matter, INFM- Roma Tre,
Via della Vasca Navale 84, 00146 Rome - Italy}
\date{\today}
\begin{abstract}
We predict that an ultra-cold Bose gas in an optical lattice
can give rise to a new form of condensation, namely matter X waves.  
These are non-spreading 3D wave-packets which reflect the symmetry of the Laplacian with a negative effective mass along the lattice direction, and are allowed to exist in the absence of any trapping potential even in the limit of non-interacting atoms. This result has also strong implications for optical propagation in 
periodic structures.
\end{abstract}
%
\maketitle
Matter waves are a natural manifestation of large scale coherence of an ensemble of atoms populating a fundamental quantum state. 
The observation of Bose-Einstein condensates (BECs) in dilute ultra-cold alkalis \cite{Anderson95Davis95} has initiated the exploration of many intriguing properties of matter waves, whose mascroscopic behavior can be successfully described via mean-field approach in terms of a single complex wave-function with well defined phase across the atom cloud \cite{Dalfovo99}.
Large scale coherence effects are usually observed by means of
3D magnetic or optical confining potentials
(also 1D cigar-like or 2D disk-shaped BEC are possible \cite{gorlitz01})  
in which BECs are described by their ground-state wave-function 
However, trapping can also occur in free space (i.e. without a trap) through the mutual compensation of the leading-order (two-body) interaction potential and kinetic energy, leading to bright (dark) solitons for negative (positive) scattering lengths. This phenomenon has been observed only
in 1D (i.e., soliton waveguides, a trap confining
the BEC transversally to the soliton) \cite{Khaykovich02}.
In 2D and 3D, free-space localization cannot occur due to
collapse instability of solitons, and even in a trap
collapse usually prevents stable formation of BEC \cite{Pitaevskii96,Berge00},
needing stabilizing mechanisms \cite{Saito03}.
\newline\indent
A lot of attention was also devoted to periodic potentials
due to optical {\em  lattices} \cite{Anderson98}, where the 
behavior of atom mimic those of electrons in crystals or photons 
in periodic media \cite{Ostrovskaya03}, and exhibit effects which stem from genuine atom coherence \cite{Burger01}. In 1D (elongated) lattices
bright (gap) solitons can form also in the presence 
of repulsive interactions \cite{Potting03}.
In this letter we predict that a novel trapping phenomenon
occurs when the full 3D dynamics is retained in a 1D lattice.
Specifically, under the conditions for which the Bloch state
associated with the lattice has a negative effective mass,
the natural state of the BEC is a localized {\it matter X wave}
characterized by a peculiar bi-conical shape. 
The atoms are organized in this way in the
absence of any trap, solely as the result of the strong anisotropy
between the 1D modulation and the free-motion in the 2D transverse plane.
Furthermore, due to axial symmetry, the atoms can experience a collective motion with given velocity along the lattice,
resulting into wave-packets traveling undistorted.
Remarkably, these matter X waves are not
only observable individually, but allow also to describe 
physically realizable BECs as their superposition, 
each component maintaining a constant (in time) number of atoms.
\newline\indent
The very first step of our derivation is similar to that 
of matter gap solitons.
Starting from the (mean-field) Gross-Pitaevskii (GP) equation \cite{Dalfovo99}
with an optical standing wave potential and no additional trap 
(we set $\eta \equiv \frac{\hbar^2}{2m}$)
\begin{equation}\label{eq:GPE}
 i \hbar \partial_{t} \psi=-\eta \nabla^2\psi+
4\Gamma \sin^2 (k z/2) \psi+a |\psi|^2 \psi=0,
\end{equation}
we assume  axial symmetry around $z$ and decompose 
the wave-function $\psi=\psi(r,z,t)$ ($r^{2} \equiv x^{2}+y^{2}$) into its forward and backward components as
\begin{equation}\label{eq:ansatznl1}
  \psi=[\psi_f(r,z,t) e^{i k z/2}+\psi_b(r,z,t) e^{-i k z/2}]
e^{i\frac{k^2-8\Gamma}{4\hbar}t},
\end{equation}
which, for $\psi_{f,b}$ slowly varying in $z$,
and dropping rapidly rotating terms, allows us to reduce
the GP equation (\ref{eq:GPE}) to the coupled equations
\begin{equation}
  \label{eq:CMTnl1} \begin{array}{l}
{\cal L}_{+}  \psi_{f} + \Gamma \psi_{b} 
- a(|\psi_f|^2 +2 |\psi_b|^2 )\psi_f =0,\\ 
{\cal L}_{-}  \psi_{b} + \Gamma \psi_{f}
-a(|\psi_b|^2 +2 |\psi_f|^2 )\psi_b =0,
\end{array}
\end{equation}
where ${\cal L}_{\pm} \equiv i \hbar \partial_{t} \pm i \eta k \partial_{z} +\eta \nabla_\perp^2$, and 
$\nabla_\perp^{2} \equiv \partial_{r}^{2}+r^{-1}\partial_{r}$.
In the linear limit $a=0$, the plane-wave [$\exp(i \kappa z-iEt/\hbar)$] linear dispersion relation associated with Eqs. (\ref{eq:CMTnl1}) has two branches $E=E_{\pm}(\kappa)=\pm \Gamma\sqrt{1+p^2}$ 
(we set $p \equiv \kappa k \eta/\Gamma$), exhibiting an energy gap of width $2 \Gamma$. The coupling between $\psi_{f}$ and $\psi_{b}$
causes the structure to be strongly dispersive near band-edge and the linear dynamics of atoms to be governed by strong Bragg reflection. 
Nevertheless, in the 1D limit ($\eta=0$), where
Eqs. (\ref{eq:CMTnl1}) were obtained previously \cite{Potting03}, 
the nonlinearity (both attractive $a<0$ and repulsive $a>0$)
induces self-transparency mediated by a two-parameter family of moving bright gap solitons, so-called because they exist in the gap seen in the soliton moving frame \cite{Conti01}. In the attractive case, one might think that the nonlinearity can balance also the kinetic transverse term $\nabla_\perp$  
leading to bell-shaped 3D atom wave-packets \cite{Aceves95}. We show in the following that, contrary to this expectation, close to the lower band edge $E=E_{-}$, the atomic wave-function takes a completely different form. 
To this end we apply a standard envelope function 
(or effective mass \cite{Meystre03}) approximation  \cite{Desterke90},
searching for spinor solutions
${\overrightarrow \psi}=[\psi_{f}~\psi_{b}]^{T}$ of the form
\begin{equation}
  \label{eq:MMS1}
  {\overrightarrow \psi}=
\epsilon \phi(\epsilon r,\epsilon z,\epsilon t)
{\overrightarrow \psi}_{-} \exp(i\kappa z-it E_{-}/\hbar)+O(\epsilon^{2})
\end{equation}
where $\epsilon$ is a small expansion parameter, 
$\phi$ is slowly modulating the
Bloch state with amplitude ${\overrightarrow \psi}_{-}=[\psi_{f-} ~\psi_{b-}]^{T}$ (eigenvector of  Eqs. (\ref{eq:CMTnl1}) with $a=0$ corresponding to the eigenvalue $E_{-}$). 
At the leading order we find that $\phi$ obeys the following asymptotic equation
\begin{equation}
  \label{eq:NLSgeneral}
i \hbar \partial_{t} \phi + i E'_{-}  \partial_{z} \phi
+ E''_{-} \partial_{z}^{2} \phi
+ \eta \nabla_\perp^2 \phi - \chi |\phi|^2 \phi=0 
\end{equation}
where $\chi=\frac{a}{2}\frac{3+2 p^2}{1+p^2}$, and
$E'_{-} \equiv \frac{dE_{-}}{d\kappa}$, 
$E''_{-} \equiv \frac{d^{2}E_{-}}{d\kappa^{2}}$
account for dispersion. For sake of simplicity we deal
henceforth with the strict band-edge case $\kappa=0$, 
which can be prepared by acting on the 
wave-number and on the potential parameter
\cite{Carusotto02}. 
In this case Eq. (\ref{eq:NLSgeneral}) reads explicitly as
\begin{equation}
  \label{eq:NLSB}
i\hbar \partial_{t} \phi
+\frac{\hbar^2}{2m} \left( \nabla_{\perp}^2 - \frac{m}{m_{e}} \partial_{z}^2 \right)\phi -\frac{3a}{2} |\phi|^2 \phi =0,
\end{equation}
where $-m_{e}\equiv -m \Gamma/(\eta k^{2})$ is the negative effective mass associated with the lattice, in turn determining the
{\em hyperbolic} character of (GP) Eq. (\ref{eq:NLSB}). 
The scaling transformation $z,r,t,\phi \rightarrow 
z_{0} z, r_{0} r, t_{0} t, c_{0} \phi$ with $r_{0}^{2}=z_{0}^{2} m_{e}/m$,
$t_{0}=2m_{e}z_{0}^{2}/\hbar$, $c_{0}^{2}=2\hbar/(3|a|t_{0})$, $z_{0}$
being a length scale, allows us to use dimensionless variables.
In the attractive case ($a<0$), Eq.  (\ref{eq:NLSB}) 
admits matter X waves solutions of the kind 
$\phi=\varphi(r,z) \exp(i \mu t)$, 
where $\varphi(r,z)$ is indeed an X-shaped invariant envelope \cite{Conti02}. 
Notice that X waves, well known in optics \cite{Saari97}, acoustics \cite{Lu92}, or microwaves \cite{Mugnai00}, as non-spreading (in space and time) solutions of the {\em linear} Helmotz wave equation \cite{Salo00},
have been only recently discovered  for Sch\"{o}dinger-type models of the form of Eq. (\ref{eq:NLSB}), by analyzing the so-called paraxial (linear \cite{Porras03} or nonlinear \cite{Conti02}) propagation in dispersive media. However, we emphasize that, while an optical or acoustic field retains a directly observable spatio-temporal X-shape \cite{Saari97,Conti02}, in the case of a Bose gas, the local density of atoms  $|\psi|^{2}$ has an X-shaped spatial envelope $|\phi|^{2}$ modulating
a term $\cos^{2}(kz/2)$ (periodic with lattice period),
due to the form of Eq. (\ref{eq:ansatznl1}).
\begin{figure}
\includegraphics[width=90mm]{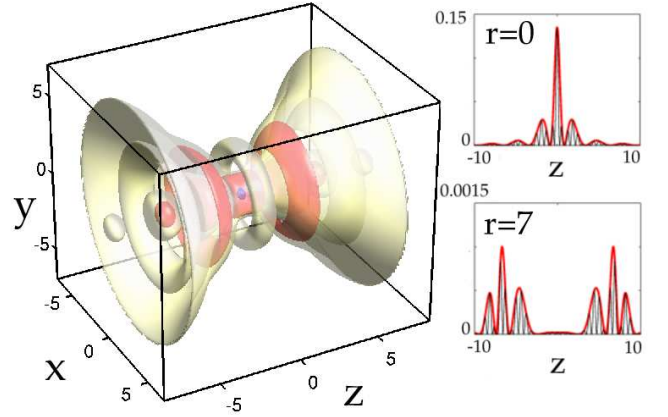}
\caption{(Color online) Surfaces of constant (high, intermediate, and low
in blue, red, and bronze, respectively)
envelope atom density $|\phi|^{2}$ 
of a matter envelope X wave solution ($\mu=0$) of Eq. (\ref{eq:NLSB}).
The insets show on-axis ($r=0$) and off-axis ($r=7$)
longitudinal profiles of the overall ($|\psi|^{2}$, black) 
and envelope ($|\phi|^{2}$, red) densities. We fix $z_{0}=10/k$
to have dimensionless units.}
\label{defocusing}\end{figure}
\newline\indent
To assess further the regimes of observability of matter X waves,
we discuss two crucial issues. 
First, by extending the analysis of Ref. \cite{Conti02}, 
we are able to show that X wave solutions of Eq.  (\ref{eq:NLSB}) 
exist also in the (far more common) case of repulsive nonlinearity ($a>0$). 
As an example, we display in Fig. \ref{defocusing} 
the atom density corresponding to the stationary 
solution with eigenvalue $\mu=0$,
which shows a dense core accompanied by bi-conically shaped
regions of lower density. As shown in the insets,
the signature of the X-shape is a single peak on-axis ($r=0$) 
and a double peak off-axis ($r\neq 0$). The envelope $|\phi|^{2}$
exhibits also slow oscillations and modulate 
the fast sinusoidal variation of the density $|\psi|^{2}$.
\newline\indent
Second, our aim is to show that 
non-spreading atom X-shaped BECs can be formed 
also in an {\em ideal noninteracting gas} ($a=0$), which is realizable by exploiting Feshbach resonance \cite{Khaykovich02}. This is in contrast with other known settings where BEC needs either a confining potential (as in pioneering experiments \cite{Anderson95Davis95}), 
or the nonlinearities to balance kinetic spreading. 
Ultimately, this stems from the fact that X waves have a finite linear limit, 
unlike solitons of standard (elliptic) GP equation whose amplitude
vanishes as $a \rightarrow 0$. To demonstrate, however, 
that {\em linear} matter X waves are observable,
we need to address the fundamental issue of their norm finiteness.
To this end let us solve  Eq. (\ref{eq:NLSB}) in dimensionless 
form with $a=0$
\begin{equation}
\label{SVEA1}
\left( i\partial_t  +\nabla^2_\perp - \partial_z^2 \right) \phi=0,
\end{equation}
by seeking for envelopes moving with velocity $v$
\begin{equation}\label{movingansatz}
\phi(r,z,t)=\varphi(r, \zeta)\exp(-i v z/2 + i t v^2/4)\text{,}
\end{equation}
where $\zeta \equiv z-v t$, and $\varphi$ turns out to obey the equation
\begin{equation}
\label{xwavereduced}
\partial_{r}^{2}  \varphi + r^{-1}\partial_{r} \varphi -\partial^2_{\zeta}\varphi=0.
\end{equation}
The general solution of Eq. (\ref{xwavereduced})
\begin{equation}
\label{Xwaveintegral}
\varphi(r,\zeta)=\int_0^\infty f(\alpha) J_0(\alpha r) e^{i \alpha \zeta} d\alpha\text{,}
\end{equation}
represents a class of envelope X waves specified
by their spectrum  $f(\alpha)$, which generalize to 
(Schr\"{o}dinger-type) Eq. (\ref{SVEA1})
the X wave solutions of Helmotz equation \cite{Salo00}. 
An exponentially decaying (with arbitrary inverse width $\Delta$)
spectrum $f_x(\alpha)=\exp(-\alpha \Delta)$, 
yields the simplest (or fundamental) X wave 
$\varphi_x=\left[ r^2+(\Delta-i\zeta)^2 \right]^{-1/2}$, while
$f_x^{(n)}=\alpha^n \exp(-\alpha \Delta)$ ($n=1,2,...$)
defines the {\it derivative} X waves $\varphi_x^{(n)}=d^n \varphi_x/d \Delta^n$.
In general from Eq. (\ref{Xwaveintegral}), 
we obtain the full envelope X wave solution of Eq. (\ref{SVEA1}) as
\begin{equation}
\phi(r,z,t)=\int_0^\infty f(\alpha) J_0(\alpha r) 
e^{i(\alpha-\frac{v}{2})z+i(\frac{v^2}{4}-\alpha v)t} d\alpha,
\end{equation}
where $|\phi|^{2}$ clearly travels undistorted along $z$. 
However, we have to face the pitfall
that these X waves do not represent physical objects since their
norm diverges. This follows from the {\it transverse scalar product}
$<\phi|\hat{\phi}>_{\perp} (z,t)$
\footnote{We define the 2D transverse scalar product 
between two  cylindrically symmetric functions 
$\phi, \eta$ as $<\phi|\eta>_\perp=2\pi
\int_0^{\infty} \phi(r,z,t) \eta^*(r,z,t) r dr$, 
as well as the 3D (in x,y,z) scalar product 
$<\phi|\eta>=\int_{-\infty}^{\infty}<\phi|\eta>_\perp dz$.} 
which yields for any pair of solutions $\phi, \hat{\phi}$ 
of Eq. (\ref{SVEA1}) with different spectra $f,\hat{f}$ (but equal velocity $v$)
\begin{equation}\label{xrproduct}
<\phi|\hat{\phi}>_{\perp}=2\pi \int_0^\infty f(\alpha) \hat{f}^*(\alpha) \alpha^{-1} d\alpha.
\end{equation}
Since $<\phi|\hat{\phi}>_{\perp}$ does not depend on $z$ and $t$, 
the 3D norm $<\phi|\phi>$ of any X wave $\phi$ diverges, 
thus requiring an (unphysical) infinite number of atoms.
Remarkably, however, finite norm beams can be generally 
constructed by introducing new orthogonal X-waves \cite{Salo01}. 
Inspired by Eq. (\ref{xrproduct}),
we can exploit the orthogonality of associated Laguerre
polynomials $L_q^{(1)}(x)$ ($q=0,1,2,...$)
with respect to the function $x \exp(-x)$,
to introduce a numerable class of (transversally) orthogonal X waves $\phi_q^{\perp}(r,z,t|v)$ defined by the 
following spectra (and parametrically by their velocity $v$) 
\begin{equation}\label{orthspectrum}
f^{\perp}_{q}(\alpha)=\frac{\Delta \alpha}{\pi \sqrt{2(q+1)}}\,L_q^{(1)}(2\Delta\alpha)
e^{-\Delta\alpha}\text{.}
\end{equation}
The X waves $\phi_q^{\perp}$ satisfy the orthogonality relation 
$<\phi_p^{\perp}|\phi_q^{\perp}>_\perp=\delta_{pq}/4 \pi$,
with $\delta_{pq}$ the Kronecker symbol. 
Importantly, the 3D scalar product shows that 
such waves are orthogonal also with
respect to the velocity, i.e. any pair of waves with
velocity $u$ and $v$ satisfies the relation 
\begin{equation}\label{vort}
<\phi^{\perp}_p(r,z,t|v)|\phi^{\perp}_q(r,z,t|u)>=
\delta_{pq} \delta(v-u).
\end{equation} 
From Eq. (\ref{vort}) it is natural to consider the  
solution $\phi=\phi_{\Sigma}$ of Eq. (\ref{SVEA1}) 
given by the superposition of orthogonal X waves 
$\{ \phi_q^{\perp}\}$ travelling with different velocities $v$ as
\begin{equation}\label{sigmabeam}
\phi_\Sigma(r,z,t)=\sum_q \int_{-\infty}^{\infty} C_q(v) \phi^{\perp}_q(r,z,t|v) dv\text{.}
\end{equation}
From the orthogonality relation (\ref{vort}), we find that
the total number of atoms is
\begin{equation} \label{norm}
\mathcal{N}_\Sigma=<\phi_\Sigma|\phi_\Sigma>=\sum_q \mathcal{N}_q,
\end{equation}
where $\mathcal{N}_q=\int_{-\infty}^{\infty} |C_q(v)|^2 dv$ represents
the atom number of the $q$-th X wave component $\phi_q^{\perp} (v)$ of the wave-packet.
Therefore we obtain the remarkable result that, while the superposition 
$\phi_\Sigma$ generally describes atom wave-packets which evolve
in time, such evolution preserves the distribution of atom number
among the X wave components.
\newline\indent
The importance of Eq. (\ref{sigmabeam}) stems from
the fact that $\phi_{\Sigma}$ describes a wide class of physical atom beams.  
To show this, we start from the integral representation of $\phi^{\perp}_q$ 
in term of its spectrum (\ref{orthspectrum}), which yields
\begin{equation}
\label{spectra1}
\phi_\Sigma=\int_{-\infty}^{\infty}\int_{0}^{\infty} F(\alpha,v)J_0(\alpha r)
e^{i(\alpha-\frac{v}{2})\zeta-i \frac{v^2}{4} t}\alpha d\alpha dv\text{,}
\end{equation}
where $F(\alpha,v)=\alpha^{-1} \sum_q C_q(v) f^{\perp}_{q}(\alpha)$.
By introducing new variables $k_t, k_z$ such that $\alpha=k_{t}$ 
and $v=2(k_t-k_z)$, and setting $U(k_t, k_z)=2 F(k_t, 2k_t-2k_z)$,
Eq. (\ref{spectra1}) can be cast in the form
\begin{eqnarray}
\label{spectra3}
\phi_\Sigma=\int_{-\infty}^{\infty} \int_{0}^{\infty}
U(k_t, k_z) J_0(k_t r)e^{i k_z z} e^{i (k_z^2-k_t^2)t} k_t dk_t dk_z,\nonumber
\end{eqnarray}
which represents the generic axisymmetric solution of Eq. (\ref{SVEA1}), 
expressed in 3D momentum space $(k_{t}, k_{z})$, 
$k_{t}$ being the momentum
transverse to the lattice direction $z$.
Here $U(k_t, k_z)$ is the Fourier-Bessel (or plane-wave) spectrum
of the initial atom distribution $\phi_0(r,z) \equiv \phi(r,z,t=0)$, 
and $U(k_{t},k_{z})=2 k_{t}^{-1} \sum_q C_q(2k_{t}-2k_{z}) f^{\perp}_{q}(k_{t})$
stands for its expansion in terms generalized Laguerre polynomials
[any square-integrable $f(x)$ in $x \in [0,\infty)$ can be expanded in terms of $L_q^{(1)}(x)$]. 
This argument can be reversed
by stating that, given the initial ($t=0$) distribution of atoms 
$U(k_t,k_z)$ in momentum space, if $F(\alpha,v)=U(\alpha,\alpha-v/2)/2$
is square integrable with respect to  $\alpha$  (with $\alpha \in [0,\infty)$), 
then the atom wave-packet admits the representation (\ref{sigmabeam}).
The expansion coefficient can be easily calculated as
\begin{eqnarray}
\label{coeffs1}
C_q(v)=\frac{\sqrt{2} \Delta}{\pi \sqrt{q+1}}
\int_{0}^{\infty}U\left(\alpha,\alpha-\frac{v}{2}\right)
e^{-2\alpha \Delta} L_q^{(1)}(2\alpha \Delta) \alpha d\alpha,\nonumber
\end{eqnarray}
Clearly, most of physically relevant wave-packets 
belong to the class $\phi_\Sigma$
\footnote{Exceptions can be found, e.g. if 
$U(k_t,k_z)=U(k_z-k_t)$, then $F(\alpha,v)$ does not depend on $\alpha$
and cannot be expanded in terms of $L_q^{(1)}$.}.
For example a spectrally narrow gaussian beam 
can be described by few X waves
(details will be given elsewhere). 
\newline\indent
Once the existence of finite norm linear X waves is established together
with their potential to describe general BECs, 
the most intriguing question remains whether we can expect matter
X-shaped atom distributions to be observable. 
Such waves correspond to a single
fixed value $q=\bar{q}$ in Eq. (\ref{sigmabeam}) and an ideal velocity distribution $C_{\bar{q}}(v)=\delta(v-\bar{v})$. 
However, more generally, we can consider an atomic envelope
beam $\phi_a$ constituted by several replicas of the single X wave 
$\phi_{\bar{q}}^{\perp}$ travelling with different velocities. 
Although, strictly speaking, such beam is not stationary, 
it can approximate such a state with an arbitrary degree of accuracy.
In other words, it is possible to construct solutions that preserve their shape,
for an arbitrary long time.  Indeed if $C_{\bar{q}}(v)$ is a
narrow function, e.g. peaked around $v=0$, Eq. (\ref{sigmabeam}),
using also Eq. (\ref{movingansatz}), 
yields the following atom envelope $\phi_a$
\begin{eqnarray}
\phi_a(r,z,t)&=& \int_{-\infty}^{\infty} \varphi^{\perp}_{\bar{q}}(r,z-vt) C_{\bar{q}}(v)
e^{-i\frac{v}{2}z+i\frac{v^2}{4}t}dv \nonumber \\ 
&\cong&  \varphi^{\perp}_{\bar{q}}(r,z)c(z,t)
\label{singleX} \end{eqnarray}
where $c(z,t)=\int_{-\infty}^{\infty}C_{\bar{q}}(v)
\exp(-i\frac{v}{2}z+i\frac{v^2}{4}t) dv$ is a solution
of the dispersive wave equation $i\partial_t c-\partial_z^2 c=0$. 
Eq. (\ref{singleX}) represents an X wave modulated by a dispersing wave,
which gives a sort of {\it adiabatic} dynamics of the finite norm
X wave. Indeed the atom beam has an invariant spatial shape fixed by $\varphi^{\perp}_{\bar{q}}$, 
which we display in Fig. \ref{fig2} as an example for $\bar{q}=0$. 
It spreads on a characteristic time which is longer, 
the narrower is the velocity distribution function $C_q(v)$.  
Finally, while in the linear regime we expect that such
atom states should be somehow prepared, we envisage that
atom collisions (nonlinear regime) can strongly favour the formation of X waves from more conventional ball-shaped atom clouds (e.g., obtained by a  harmonic 3D trap which is then switched off) 
through instability mechanisms \cite{Conti02,Conti03}, 
an issue which will be deepened elsewhere.
\begin{figure}
\includegraphics[width=90mm]{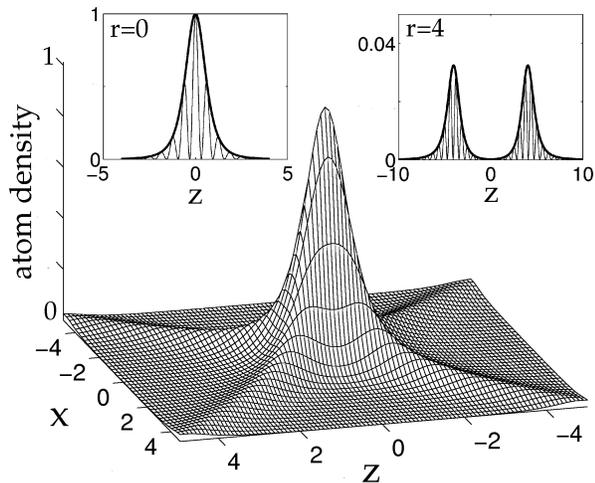}
\caption{Atom density of a pure X wave
$|\phi^{\perp}_{q}(r,z)|^2$ with $q=0$ and $\Delta=1$,
as seen on the $(x,z)$ plane ($y=0$).
The insets show (same units as in Fig. 1)
the overall (thin line) and envelope
(bold line) density on-axis and off-axis, respectively.}
\label{fig2} \end{figure}
\newline\indent
In conclusion we have shown that a periodic potential
supports moving or still localized states of the GP model with 
envelope X-shape (strictly or nearly) preserved upon motion. 
A matter X waves entails localization both in momentum and configuration space and is a clear signature of a Bose condensed gas, so much as the anisotropy in the distribution function \cite{Dalfovo99} detected in early experiments.
However, unlike any other form of BEC including solitons, 
matter X waves can be observed in free-space and in the
non-interacting regime, where they constitute a natural basis of expansion to describe the coherent properties of atom wave-packets. 
These results have strong implications also in optics, 
where the model (\ref{SVEA1}) holds for normally dispersive bulk media \cite{Porras03}, 
or in nonlinear optics of stratified media 
where Eqs. (\ref{eq:CMTnl1}) describe propagation in a 1D
bulk grating in the presence of diffraction, or propagation
in a fully 3D photonic crystal (along proper directions).
\begin{acknowledgments}
We thank F. Cataliotti, M. Oberthaler, M. Salerno and P. Di Trapani for discussions.
C.C. thanks the Fondazione Tronchetti Provera for the financial support. 
\end{acknowledgments}


\begin{thebibliography}{30}

\bibitem{Anderson95Davis95}
M.~H. Anderson {\it et al.},
Science {\bf 269}, 198 (1995);
K.~B. Davis {\it et al.},
 \prl {\bf 75}, 3969 (1995).

\bibitem{Dalfovo99}
F.~Dalfovo, S.~Giorgini, L.~Pitaevskii,
and S.~Stringari, \rmp {\bf 71}, 463 (1999).

\bibitem{gorlitz01}
A.~Gorlitz {\it et al.},
\prl {\bf 87}, 130402 (2001).

\bibitem{Khaykovich02}
L.~Khaykovich {\em et al.}, 
Science \textbf{296}, 1290 (2002);
K.~Strecker, G.~Partrige, A.~Truscott, and
R.~Hulet, Nature (London) \textbf{417}, 150 (2002).

\bibitem{Pitaevskii96}
L.~Pitaevskii, Phys. Lett. A  \textbf{221}, 14 (1996).

\bibitem{Berge00}
L.~Berge, T.~J. Alexander, and Y.~S. Kivshar, 
\pra \textbf{62}, 23607 (2000).

\bibitem{Saito03}
H.~Saito and M.~Ueda, \prl 
\textbf{90}, 40403
  (2003).

\bibitem{Anderson98}
B.~P.Anderson and M.~Kasevich, Science \textbf{282},
1686 (1998).

\bibitem{Ostrovskaya03}
E.A. Ostrovskaya and Y.S. Kivshar,
\prl {\bf 90}, 160407 (2003).

\bibitem{Burger01}
S.~Burger {\em et al.}, 
\prl  \textbf{86}, 4447 (2001);
O.~Morsch, J.~H.Muller, M.~Cristiani, D.~Ciampini, and E.~Arimondo,
\prl \textbf{87}, 140402 (2001);
A.~Trombettoni and A.~Smerzi,
\prl \textbf{86}, 2353 (2001);
F.~S.Cataliotti {\em et al.},
Science \textbf{293}, 843 (2001);
Greiner {\em et al.}, Nature (London) \textbf{419}, 51 (2002);
B.~Wu and Q.~Niu,
\pra \textbf{64}, R061603 (2001);
H.~P.Buchler, G.~Blatter, and W.~Zwerger,
\prl \textbf{90}, 130401 (2003).

\bibitem{Potting03}
S.~P\"{o}tting, P.~Meystre, and E.~M.Wright, in
\emph{Nonlinear photonic crystals}
(Springer, 2003), vol.~10 of \emph{Photonics}.

\bibitem{Conti01}
C.~Conti and S.~Trillo,
\pre \textbf{64}, 036617 (2001).

\bibitem{Aceves95}
A.~Aceves, B.~Costantini, and C.~De Angelis, 
\josab \textbf{12}, 1475 (1995).

\bibitem{Desterke90}
C.~M. de~Sterke and J.~E. Sipe,
\pra \textbf{42}, 550  (1990).

\bibitem{Meystre03}
H. Pu {\it et al.},
\pra \textbf{67}, 043605  (2003).

\bibitem{Carusotto02}
I. Carusotto, D. Embriaco, and G. C. LaRocca,
\pra \textbf{65}, R053611 (2002).

\bibitem{Conti02}
C.~Conti {\em et al.}, 
\prl {\bf 90}, 170406 (2003); see also P.~D.Trapani {\em et. al.},  
physics/0303083 (2003).

\bibitem{Saari97} 
P.~Saari and K.~Reivelt,
\prl \textbf{79}, 4135 (1997).

\bibitem{Lu92}
J.~Lu and J.~F.Greenleaf,
IEEE Trans. Ultrason. Ferrelec. Freq. contr.
 \textbf{39}, 441 (1992).

\bibitem{Mugnai00}
D.~Mugnai, A.~Ranfagni, and
R.~Ruggeri, \prl \textbf{84}, 4830 (2000).

\bibitem{Salo00}
J.~Salo, J.~Fagerholm, A.~T.Friberg,
and M.~M. Salomaa, \pre \textbf{62}, 4261 (2000),
and references therein.

\bibitem{Porras03}
M.~A.Porras, S.~Trillo, C.~Conti, and P.~D.Trapani, 
\ol \textbf{28}, 1090 (2003).

\bibitem{Conti03}
C.~Conti, physics/0302053; C.~Conti, \pre \textbf{68}, 016606 (2003).

\bibitem{Salo01}
J. Salo and M. M. Salomaa, J. Phys. A: Math. Gen.,
{\bf 34}, 9319 (2001).

\end{thebibliography}
\end{document}